\documentclass[twocolumn,showpacs,amsmath,amssymb,prl]{revtex4}

\usepackage{graphicx}
\usepackage{bm}

\begin{document}

\title{Intervortex quasiparticle tunneling and
electronic structure of multi-vortex configurations in type-II superconductors.}

\author{A.\,S.\,Mel'nikov,
M.\,A.\,Silaev}

\address{Institute for Physics of Microstructures, Russian Academy of Sciences,
603950, Nizhny Novgorod, GSP-105, Russia}

\date{\today}

\begin{abstract}
The electronic spectrum of multi-vortex configurations in type-II superconductors is
studied taking account of the effect of quasiparticle tunneling between the vortex cores.
 The tunneling is responsible for the formation
  of strongly coupled quasiparticle states for intervortex distances
  $a<a_c$, where the critical distance $a_c$ is
 of the order of several coherence lengths $\xi$.
Analyzing the resulting spectra of vortex clusters bonded by the quasiparticle tunneling
we find a transition  from a set of
degenerate Caroli - de Gennes - Matricon branches to anomalous branches similar
to the ones in multi-quanta giant vortices.
This spectrum transformation  results in the oscillatory
behavior of the density of states at the Fermi level
as a function of $a$ and could be observed
in mesoscopic superconductors and disordered flux line arrays in the bulk systems.
\end{abstract}


 \maketitle

Theoretical investigations of the quantum mechanics of low energy
quasiparticle (QP) excitations in isolated vortices and vortex arrays
are of crucial importance for understanding the basic thermodynamic and transport
phenomena in the mixed state of type-II superconductors at low temperatures.
This issue has been studied for several decades starting from the
pioneering work by Caroli, de Gennes and Matricon (CdGM) \cite{CdGM}.
For an isolated singly-quantized vortex
there appears a so-called anomalous branch of the subgap spectrum
which corresponds to the QP states bound
to the vortex core because of the Andreev reflections from the gap potential profile
$\Delta = |\Delta(r)|e^{i\theta}$, where $(r,\theta,z)$ is a cylindrical coordinate
system, and $z-$ axis is chosen parallel to the vortex line.
The low energy spectrum of these localized states can be written as follows:
$E=-\mu\omega\simeq -\mu \Delta_0/(k_\perp\xi)$,
where $\Delta_0$ is the gap value far from the vortex
core, $\xi$ is the superconducting coherence length, $k_\perp =\sqrt{k_F^2-k_z^2}$
is the Fermi momentum projection on the plane perpendicular to the vortex axis,
and the angular momentum quantum number $\mu$ is half an odd integer.
The overlapping of QP wave functions of neighboring vortices should
perturb the CdGM spectrum and result, e.g., in the
band structure effects in periodic vortex lattices \cite{bands}.
With the increase in the vortex concentration at high magnetic fields close to the upper
critical field $H_{c2}$ there occurs a crossover to the Landau-type quantization for QPs
precessing along the whole Fermi surface (see \cite{dHvA} and references therein).
A scenario of such crossover from discrete CdGM levels to the Landau-type spectrum
is an appealing problem which would allow to understand the nature of de Haas - van
Alphen oscillations
observed experimentally \cite{exp} even for magnetic fields  $H\sim (0.3-0.4) H_{c2}$
when the vortices are well-separated.
The transformation of the QP spectrum in the latter case should be controlled
by the effect of QP tunneling between the vortex cores.
This tunneling phenomenon is expected to play an essential role
also for exotic vortex configurations formed in mesoscopic superconductors
of the size of several coherence lengths.
In such systems the balance of competitive forces acting on vortices due to the screening current
flowing at the sample boundary and the intervortex repulsion,
leads to the formation of small size multi-vortex configurations
(vortex molecules)  and multi-quanta (giant) vortices \cite{Geim}.
These vortex states can transform into each other via magnetic field-driven
first or second order phase transitions.
According to the general theory \cite{volovik1} the number of anomalous branches (per spin)
crossing the Fermi level
at certain impact parameters $\mu_F/ k_\perp\lesssim \xi$
equals to the winding number $M$ of a multi-quanta vortex.
The anomalous branch with $\mu_F=0$ responsible for the peak in
the density of states (DOS) at the vortex center
exists only for vortices with an odd winding number.
 Generally the spatial distribution of the
DOS has the shape of rings with radii of the order of
$\xi$ \cite{multi,MV}.
The splitting of a multi-quanta vortex into $M-$ vortex molecule
with the intervortex distance $a$ results in the transformation of these rings in the DOS profile
into a set of peaks at the centers
of individual vortices. The initial stage of such DOS transformation has been studied
in \cite{MV} by the perturbation method valid for $a\ll\xi$.

In this Letter we address the limit $a>\xi$ of well-separated vortices in
mesoscopic superconductors and disordered vortex arrays
in the bulk systems well below $H_{c2}$.
The goal of the present work is to study a scenario of the QP spectrum transformation
caused by the formation of vortex clusters bonded by the QP tunneling
between the Andreev wells in the cores.
Note that hereafter we neglect
all the possible normal scattering effects at impurities \cite{LarkOvch98} or
mesoscopic sample boundaries \cite{MKVR} which also can affect the low energy
spectrum.

\begin{figure}[t]
\centerline{\includegraphics[width=0.65\linewidth]{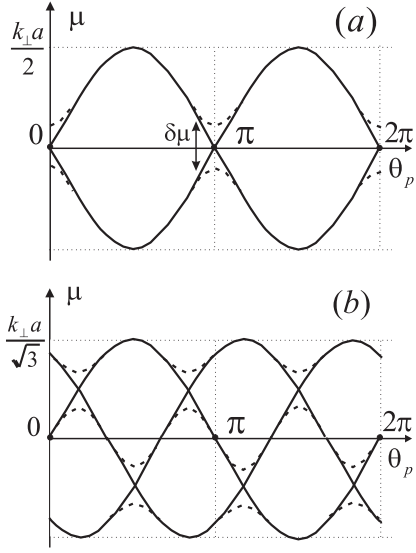}}
\caption{Fig.1. Angular momentum $\mu$ as a function of $\theta_p$ at $E=0$ for (a)
two  and (b) three vortices
}
 \label{fig1}
\end{figure}

Let us start with a qualitative analysis of the intervortex tunneling effect
and consider a set of vortex lines parallel to the $z-$ axis. In the plane $(xy)$
the vortex centers defined as zeros of the superconducting order
parameter are positioned at certain points ${\bf r}_i$.
For QPs propagating along the classical trajectories parallel to
${\bf k}_\perp =k_\perp(\cos\theta_p,\sin\theta_p)$
we introduce the angular momenta
 $\mu=[{\bf r},{\bf k}_\perp]\cdot{\bf z_0}=k_\perp r\sin(\theta_p-\theta)$
 and $\tilde\mu_i=\mu-[{\bf  r}_i,{\bf k}_\perp]\cdot{\bf z}_0$
 defined with respect to the $z-$ axis
passing through the origin and with respect to the $i$-th vortex axis
passing through the point ${\bf r}_i$, correspondingly.
Neglecting the QP tunneling  between the vortex cores we get
degenerate  CdGM energy branches:
$E_i=-\omega\tilde\mu_i$.
For a fixed energy $E$ we can define a set of
crossing branches on the plane $(\mu ,\theta_p)$:
$\mu_i(\theta_p) = -E/\omega
+[{\bf r}_i,{\bf k}_\perp]\cdot{ \bf z_0}$.
These branches are shown in Fig.\ref{fig1} for
two vortices with ${\bf r}_1 = (-a/2,0)$ and ${\bf r}_2 = (a/2,0)$
and three vortices
 at the apexes of the equilateral triangle with the center at the
origin:
 ${\bf r}_1 = (0,a/\sqrt 3)$, ${\bf r}_2 = (a/2,-a/(2\sqrt 3))$
 and ${\bf r}_3 = (-a/2,-a/(2\sqrt 3))$.
Each crossing point of branches $\mu_i(\theta_p)$ and  $\mu_j(\theta_p)$
 corresponds to the trajectories
passing through the cores of $i-$th and $j-$th vortices.
It is natural to expect that the degeneracy at these points will be removed if we take account of
a finite probability of QP tunneling between the cores. The tunneling
is determined by the exponentially small overlapping of wave functions
localized near the cores and
results in the splitting of energy levels:
$\delta E\sim \Delta_0\exp(-k_F a_{ij}/(k_\perp\xi))$,
where $a_{ij}=|{\bf r}_i-{\bf r}_j|$.
The estimate for the
branch splitting
$\delta\mu$ shown in Fig.\ref{fig1} by dash lines reads:
\begin{equation}
\delta \mu(a_{ij}) \sim \frac{\delta E}{\omega}=
k_\perp\xi\exp\left(-\frac{k_F
a_{ij}}{k_\perp\xi}\right) \ .
\end{equation}
 As a result, we get the  branches $\mu^*_i(\theta_p)$ with a qualitatively
new behavior: each of these branches consists of parts corresponding to
the classical QP trajectories passing through the cores of different vortices.
Keeping in mind general conditions of the semiclassical approach
validity it is natural to expect that a reasonable criterion
on the intervortex tunneling efficiency can be obtained if we compare the
splitting $\delta\mu(a_{ij})$ with the quantum mechanical
 uncertainty $\Delta\mu$ of the angular momentum.
The latter value can be estimated from the
uncertainty principle $\Delta\mu\Delta\theta_p \sim 1$,
where the expression for the angle uncertainty near the branch crossing points
reads: $\Delta\theta_p\sim \Delta\mu/(k_\perp a_{ij})$.
Provided the branch splitting is rather small, i.e.
$\delta\mu(a_{ij})\ll\Delta\mu\sim\sqrt{k_\perp a_{ij}}$ for all crossing points,
the branches $\mu_i(\theta_p)$ are almost independent.
 In the opposite limit
$\delta\mu(a_{ij})\gg\sqrt{k_\perp a_{ij}}$
the QP states of the $i-$th and $j-$th vortices appear to be strongly
coupled by tunneling: near the branch crossing point
the QP trajectory performing a precession in the
core of the $i-$th vortex experiences a transition to the core of the
$j-$th vortex. According to the above condition on $\delta\mu(a_{ij})$
the tunneling is most efficient for $k_\perp=k_F$
and $a_{ij}<a_c$, where $a_c \simeq (\xi/2) \ln(k_F\xi)$
 is a critical intervortex distance.
Using the percolation theory language we can consider the vortices
to be bonded if $a_{ij}<a_c$ and define a cluster in a disordered
flux line system  as a set of $M$ vortices
bonded either directly or via other vortices.
The number $M$ grows
with the increase in the vortex concentration
and becomes infinite at the percolation threshold
or for a periodic vortex lattice.
Certainly in mesoscopic superconductors
the cluster dimensions $L_v$ can not exceed the sample size $R$.
Our further consideration of the bonded QP states
 is restricted to the limit
of finite clusters with $L_v\ll r_L$, where $r_L\sim k_\perp \phi_0/H$
 is the cyclotron orbit radius, and $\phi_0=\pi\hbar c/|e|$ is the flux quantum.
The cluster is characterized by a set of
hybridized QP states: with a change in the ${\bf k}_\perp$
direction the wave function experiences a number of
subsequent transitions between the cores of neighboring vortices.
Taking, e.g., the upper branch in Fig.\ref{fig1}(a)
we obtain the wave function concentrated near the cores of the right and left
vortices for the angular intervals $0<\theta_p<\pi$
and $\pi<\theta_p<2\pi$, respectively.
Further decrease in the intervortex distance results
in the increase in the tunneling probability
and, thus, the increase in $\delta\mu (a_{ij})$.
 Finally, for $a_{ij}\rightarrow 0$ we get a set of $M$ lines
$\mu ={\rm const}$ parallel to the $\theta_p$ axis, i.e. $M$ anomalous branches
crossing zero energy at angular independent impact parameters and corresponding to
the $M-$ quanta vortex. Certainly this limit can be realized only in mesoscopic
samples.

Following \cite{k-v} one can obtain the discrete energy levels applying
the Bohr-Sommerfeld quantization rule for canonically conjugate
variables $\mu$ and $\theta_p$:
\begin{equation}
\label{bohr}
  \int_0^{2\pi n_\theta}\mu(\theta_p) d\theta_p=2\pi(n+\beta),
\end{equation}
where $n$ and $1\leq n_\theta\leq M$ are integers,
$2\pi n_\theta$ is the period of the $\mu(\theta_p)$ function,
and $\beta=1/2$ ($\beta =0$) for odd (even) $n_\theta$ values.
Depending on the ratio $\delta\mu(a_{ij})/\Delta\mu$
one should apply this quantization rule
either to the branches $\mu_i (\theta_p)$ or to the branches $\mu^*_i
(\theta_p)$.
In the momentum region
\begin{equation}
k_F\sqrt{1-({\rm min}(a_{ij})/a_c)^2} \ll |k_z|  <k_F
\end{equation}
one can neglect the branch splitting
$\delta\mu(a_{ij})\ll \Delta\mu$ at all and Eq.(\ref{bohr})
written for the branches  $\mu_i (\theta_p)$ gives us the CdGM spectrum
with a minigap $\omega_0/2=\omega (k_z =0)/2$.
For ${\rm min}(a_{ij})>a_c$  the CdGM expression holds for the entire momentum range.
For vortices forming
 a cluster  the  QP states bonded by intervortex tunneling
 appear in a finite momentum interval
\begin{equation}
 |k_z|\ll k_F\sqrt{1-({\rm min}(a_{ij})/a_c)^2} <k_F \ .
\end{equation}
In this limit the QP tunneling between the cores results in the
qualitative modification of
spectrum which can be obtained by substituting $\mu^*_i(\theta_p)$
into Eq.(\ref{bohr}):
\begin{equation}
\label{double}
  E_{ni} (k_z)\approx\frac{\Delta_0}{\xi}
  \left(\frac{n+\beta}{k_\perp}+
  b_i ({\bf r}_1,..{\bf r}_M) \right) \ ,
\end{equation}
where $i=1 .. M$. The spectrum (\ref{double}) is
similar to the one of a multi-quanta
vortex \cite{volovik1,multi,MV} which recovers
in the limit $a_{ij}\rightarrow 0$ when $|b_i|\lesssim \xi$.
The multi-vortex cluster geometry
and its dimensions $L_v$ determine the effective
impact parameters $b_i ({\bf r}_1,..{\bf r}_M)$
which vary in the range $-L_v\lesssim b_i\lesssim L_v$.
Taking a two- (three-) vortex molecule with $\xi < a < a_c$
as an example we get
 $b_{1,2}\sim \pm  a$ ($b_{1,3}\sim \pm a$, $b_2=0$).
Contrary to the CdGM  case the spectrum branches
(\ref{double}) can cross the Fermi level as we
decrease the characteristic
 intervortex distance $a$ and the  minigap is suppressed.
The DOS consists of $M$ sets of van Hove singularities corresponding
to the extrema of $ E_{ni} (k_z)$ branches.
 The energy interval between the peaks belonging to each set is $\omega_0$.
For a certain fixed energy the DOS as a function of $a$
 exhibits oscillations with the period
 of the order of the atomic length scale $\delta a\sim k_F^{-1}$.
Experimentally the intervortex distance can be controlled by
a varying  magnetic field.
For typical values $a\sim\sqrt{\phi_0/H}$ we get
the following field scale of DOS oscillations:
$\delta H/H\sim \sqrt{\hbar\omega_H/\varepsilon_F}$, where $\omega_H=|e|H/mc$ is the cyclotron
frequency, $m$ is the electron effective mass, and $\varepsilon_F$ is the Fermi
energy.
The oscillatory behavior should affect both thermodynamic and transport properties
at low temperatures though in real experimental conditions the DOS peak structure is
certainly smeared due to the various mechanisms of level broadening, e.g.,
finite temperature, fluctuations in vortex positions,
impurity scattering effects, etc.
It should be noted that for typical values $k_F\xi=10^2 - 10^3$
the critical distance $a_c/\xi\sim 2-3$
exceeds the core radius and
the spectrum transformation starts
at the fields
$H\sim \phi_0/a_c^2\sim H_{c2}[{\rm ln}(k_F\xi)]^{-2}$
when the vortices are indeed well-separated.

Now we proceed with a quantitative analysis of the intervortex tunneling
on the basis of the Bogolubov - de Gennes theory:
\begin{equation}
\hat H_0\hat\Psi
+\hat\sigma_x  {\rm Re}\Delta\hat\Psi-
\hat\sigma_y  {\rm Im}\Delta\hat\Psi =E\hat\Psi \ ,
\end{equation}
where
$\hat H_0=\hat\sigma_z(\left(\hat {\bf p}-
e\hat\sigma_z{\bf A}/c\right)^2-\hbar^2k_{\perp}^2 )/2m$,
$\hat\Psi=(U,V)$,
$U$ and $V$ are the particle- and hole - like parts of the QP wave function,
 $\hat\sigma_i$ are Pauli matrices,
$\hat {\bf p}=-i\hbar\nabla$, $\Delta=\Delta(x,y)$ is an order parameter profile.
For extreme type-II superconductors we can assume
the magnetic field ${\bf H}=-H{\bf z}_0$  to be homogeneous
on the spatial scale $L_v$ and take the gauge ${\bf A}=[{\bf H},{\bf r}]/2$.
Within the quasiclassical approach the wave function in the momentum
representation can be taken in the form:
 \begin{equation}
 \label{st}
 \hat\Psi({\bf p})=
 \frac{1}{k_\perp}\int\limits_{-\infty}^{+\infty} ds
 e^{-i(|{\bf p}|-\hbar k_\perp)s/\hbar}
\hat\psi(s,\theta_p) \ .
\end{equation}
The equation for $\hat\psi(s,\theta_p)$ reads:
$\hat H\hat\psi =E \hat\psi$, where
\begin{equation}
\label{bdg}
\hat H = -i\hbar V_\perp \hat\sigma_z\frac{\partial}{\partial s}+
\hat\sigma_x  {\rm Re}\Delta-
\hat\sigma_y  {\rm Im}\Delta
 +\frac{\hbar \omega_H}{2}
\hat\mu \ ,
\end{equation}
$mV_\perp=\hbar k_\perp$,
$\Delta=\Delta(\hat x, \hat y)$,
$\hat\mu =-i\partial/\partial \theta_p$,
 and
 the terms quadratic in $H$ are neglected assuming $L_v\ll r_L$.
The coordinate operator takes the form:
\begin{equation}
\hat {\bf r} =s {\bf k}_\perp/k_\perp
 +
 \left\{ [{\bf k}_\perp, {\bf z}_0], \hat\mu\right\}/(2k_\perp^2)
\end{equation}
where $\{A,B\}=AB+BA$ is an anticommutator.
Let us emphasize that the Hamiltonian (\ref{bdg}) takes account
of noncommutability of $\hat\mu$ and $\theta_p$ and, thus,
the above description involves the angular momentum quantization.
Replacing $\hat\mu$ by a classical variable
we get Andreev equations along straight trajectories.
For an isolated vortex positioned at ${\bf r}=0$ and described by
the gap function $\Delta_1({\bf r})= \Delta_v(r)e^{i\theta}$
a standard  solution of the above equations
corresponding to low energy CdGM levels reads:
$\hat\psi=\exp (i\mu\theta_p)\hat\psi_0 (s,\theta_p)$, where
\begin{equation*}
\hat\psi_0 (s,\theta_p)=
\exp\left(
\frac{i}{2}\hat\sigma_z \theta_p
\right)
\sqrt{\frac{\Delta_0}{2\hbar V_\perp\Lambda}}
 \left(1\atop -i\right)
e^{-K(s)} \ ,
\end{equation*}
$$
K(s)=\frac{1}{\hbar V_\perp}\int\limits_0^{s}\quad
 \frac{t \Delta_v \left(t\right)}{|t|} dt \ ,
  \,
  \Lambda= \frac{2\Delta_0}{\hbar V_\perp}\int\limits_{0}^{+\infty}
  e^{-2K(t)} d t \ .
$$

The physical picture of the intervortex QP tunneling effect can be illustrated
by consideration of the simplest generic problem:
two neighboring vortices positioned at ${\bf r}_\pm = (\pm a/2,0)$.
The distances from this vortex pair to other vortices
or to a mesoscopic sample boundary are assumed to be larger
than the critical distance $a_c$ so that
their influence can be neglected.
 To describe the case of well-separated vortices $a>\xi$
 it is natural to use the tight binding
 approximation for the QP wave function:
\begin{equation}
\label{tb}
\hat \psi = c_+(\theta_p)
\widehat{T}_+ \hat\psi_0 (s,\theta_p) +
c_-(\theta_p) \widehat{T}_-\hat\psi_0 (s,\theta_p) \ ,
\end{equation}
where the operator
$$
\widehat{T}_\pm=\exp\left(\frac{i\hat\sigma_z \varphi_\pm}{2}
\mp   \frac{a\cos\theta_p}{2} \left(i k_\perp+\frac{\partial}{\partial s}
\right)\pm \frac{ik_\perp a}{2}
\right)
$$
transforms the CdGM states of a single vortex at the origin into
 the eigenstates for isolated vortices
at ${\bf r}_\pm = (\pm a/2,0)$, $\varphi_\pm$ is the order parameter phase
induced at the center of right (left) vortex due to the presence of neighboring vortices
or boundaries.
The coefficients $c_\pm$ satisfy the condition $c_\pm (\theta_p+2\pi) = -c_\pm (\theta_p)$,
so that the total wave function is single valued. In the angular intervals
$|\sin\theta_p|>\xi/a$ the QP trajectories can not pass through both vortex cores and
we have two solutions with either $c_+=0$ or $c_-=0$, i.e. the states
corresponding to one of the vortices.
The only effect of neighboring vortices or
boundaries on the QP energy for such trajectories
could be associated with the Doppler shift $\epsilon_d$
caused by the superfluid velocity field
 induced by the external sources in a given vortex core.
However this superflow would produce the Lorentz force acting on a vortex and,
thus, $\epsilon_d$ should vanish for a static vortex configuration.
This condition gives us the relation $a=a(H)$ which depends on the vortex arrangement and
sample geometry (see, e.g., \cite{Buzdin}).

\begin{figure}[t]
\centerline{\includegraphics[width=1.0\linewidth]{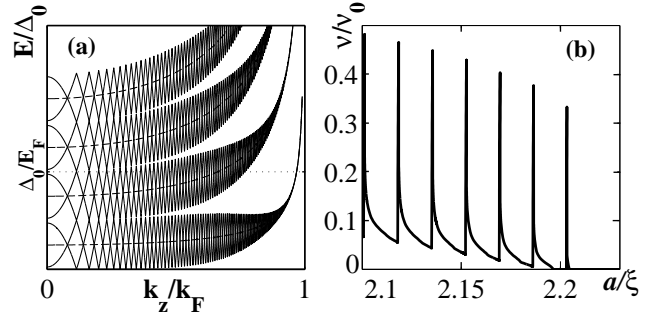}}
\caption{Fig.2.
Two-vortex molecule ($k_F\xi=200$): (a)
energy spectrum for $a=2.1\xi$; (b)
the DOS per spin projection averaged over the interval
 $\delta E= 0.05\omega_0$ around $E=0$
 as a function of the intervortex
distance $a$.
  The CdGM spectrum is shown by the dash lines.
  $\nu_0=k_F/(4\omega_0)$ is the averaged CdGM DOS
 }
 \label{fig2}
\end{figure}

The angular intervals $|\sin\theta_p|<\xi/a$ should be considered separately
since the corresponding QP trajectories pass through both vortex cores.
Substituting Eq.(\ref{tb}) into the quasiclassical equations with the Hamiltonian
(\ref{bdg}),
multiplying them by the  functions
$\left(\widehat{T}_\pm\hat\psi_0\right)^*$,
 and integrating over s, one obtains:
\begin{equation}
\label{eq-tunnel}
- \omega \hat\mu \hat c
 -t
e^{i\hat\sigma_zk_\perp a(\cos\theta_p-1)}\hat\sigma_x
\hat c =E \hat c \ ,
\end{equation}
where $\hat c =(c_+,c_-)$.
For the trajectories with $\theta_p$ close to $0$ or $\pi$ we find an approximate
expression for the intervortex tunneling amplitude:
$t\simeq \Delta_0 e^{-ak_F/(\xi k_\perp)}$.
Note that here we omit small $\theta_p-$ dependent corrections to $\omega$
resulting from the angular dependent distortions of the gap profiles in the cores
and neglect the renormalization of the $\omega$ and
$t$ values caused by the terms $\propto\hbar\omega_H\sim \omega_0(\xi/a)^2\ll \omega_0$.
Let us introduce the function
$\hat b = \exp(iE\theta_p/\omega- i k_\perp a \hat\sigma_z(\cos\theta_p-1)/2) \hat c$.
For $|\theta_p|\ll \xi/a$ the Eq.(\ref{eq-tunnel}) reads:
\begin{equation}
-\omega\hat\mu  \hat b
+\frac{k_\perp a \omega}{2} \theta_p \hat\sigma_z \hat b
-t \hat\sigma_x \hat b =0 \ .
\end{equation}
This problem is equivalent to
the one describing the interband tunneling
\cite{kane} or the one-dimensional
(1D) motion of a Dirac particle
in a uniform electric field
 and the solution can be written in terms of
  the parabolic cylinder functions $D_{i\alpha} (\sigma)$:
\begin{equation}\label{pc}
\hat b = \left( d_1 D_{i\alpha}(\sigma) + d_2 D_{i\alpha}(-\sigma)\atop
\sqrt{-i\alpha}(d_1 D_{i\alpha-1}(\sigma) - d_2 D_{i\alpha-1}(-\sigma)) \right) \ ,
\end{equation}
where $\alpha = t^2/(\omega^2k_\perp a)\sim(\delta\mu (a)/\Delta\mu)^2$, and
 $\sigma = \theta_p\sqrt{k_\perp a/i}$.
 Taking the asymptotical expressions for $D_{i\alpha} (\sigma)$
in the interval
 $ {\rm max}[\alpha,1]/\sqrt{k_\perp a}\ll|\theta_p|<\xi/a$
 we get:
 $\hat b(\theta_p>0)= \hat S \hat b(\theta_p<0)$, where
$$
\hat S =  e^{-\pi\alpha}\hat I+
ie^{i\hat\sigma_z\gamma} \left(
 \hat\sigma_y{\rm Re}\tau
+\hat\sigma_x{\rm Im}\tau
 \right)e^{-i\hat\sigma_z\gamma} \ ,
$$
$$
\tau=\tau(\alpha)= \frac{2\Gamma (1-i\alpha)}{\sqrt{-2\pi i\alpha}} e^{-\pi\alpha/2} \sinh (\pi\alpha) \ ,
$$
$\hat I$ is the unity matrix,
 $\gamma=k_\perp a\theta_p^2/4+\alpha \ln(|\theta_p|\sqrt{k_\perp a})$, and $\Gamma (x)$ is
the gamma function.
For the angles $\theta_p$ close to $\pi$ the solution can be obtained
using a transformation:
$\hat b (|\theta_p - \pi|\ll \xi/a) =\hat\sigma_x \hat b (|\theta_p|\ll \xi/a)$.
Matching the wave function in various angular domains
and using the condition $\hat c(\theta_p)=-\hat c(\theta_p+2\pi)$
we obtain the  spectrum:
\begin{equation}
\cos (\pi E/\omega)=
\pm e^{-\pi\alpha/2} \sqrt{2\sinh(\pi\alpha)}\sin(\pi\chi) \ ,
\label{spectrum}
\end{equation}
where
$\pi\chi =  k_\perp a +\alpha \ln(k_\perp\xi^2/a)+\arg(\Gamma(1-i\alpha))+\pi/4$.
The generalization of such matching procedure for  $M-$ vortex clusters
is straightforward.
The spectrum calculated using the Eq.(\ref{spectrum}) is shown in Fig.\ref{fig2}(a)
for a typical parameter $k_F\xi=200$ and for a
rather small intervortex distance $a<a_c$.
One can see that the transformation of the spectrum $E(k_z)$
occurs according to the scenario suggested above:
as we decrease the distance $a$ below $a_c$
the crossover to the double quanta
vortex spectrum starts in the region of small $k_z$ values
defined by the condition
$\alpha > 1$.
For $\alpha\ll 1$ we get the CdGM spectrum with a small oscillatory correction:
\begin{equation}
E -\omega (n+1/2) \simeq\mp (-1)^n \omega \sqrt{2\alpha/\pi}
\sin(k_\perp a +\pi/4) \ .
\end{equation}
The effective minigap
$E_{min}=\omega_0 (1/2 - \sqrt{2\alpha/\pi})$
vanishes for $a\sim a_c$.
For intermediate values $\sqrt{k_\perp \xi^2/a}\gg\alpha\gg 1$ the
spectrum takes the form (\ref{double}) if we put $b_{1,2}=\pm\chi/k_\perp$.
This expression can be matched with the one obtained
 using a standard WKB expansion for Eq.(\ref{eq-tunnel})
in the limit $\alpha\gg 1$ if we put
\begin{equation}
\frac{\pi\chi}{k_\perp a} =1 +
\int\limits_{0}^{\xi/a}\left(
\sqrt{\theta_p^2
+\frac{4\alpha}{k_\perp a}} -\theta_p
\right)d\theta_p \ .
\end{equation}
In agreement with general arguments presented above
the DOS in the limit $a<a_c$  consists of two sets of peaks
shifted by the value $\omega_0 (2\chi-[ 2\chi])$, where
$[...]$ denotes the integer part.
The oscillatory behavior of the DOS at the Fermi level
as a function of $a$ is shown in Fig.\ref{fig2}(b).

In conclusion we note that
the bending of vortex lines in a cluster can strongly affect
the behavior of the wave functions along the magnetic field direction.
Let us take, e.g., a two-vortex molecule and
assume the function $a(z)$ to change slowly on a scale
$\sim k_F^{-1}$.
One can replace the momentum $k_z$ in Eq.(\ref{double})
by the operator $\hat k_z =-i\partial/\partial z$
and consider the spectrum branches
as a set of effective Hamiltonians $E_{ni} (\hat k_z)$
describing the 1D quantum mechanics of QPs.
Such approach is analogous to the one suggested in \cite{Shytov}
for QPs in superconducting/normal metal layered structures.
Taking a simple model profile $a (z)=\bar a -\tilde a z^2/L^2$
and choosing $n\simeq \mp k_F \bar a/\pi$
such 1D problem for energies close to the $E_{ni}(k_z)$ extremum
 can be reduced to the one of
a quantum mechanical harmonic oscillator with the effective mass
$m^*\sim m k_F\xi^2/\bar a$ and interlevel spacing
$\Omega\sim \omega_0\sqrt{\bar a \tilde a}/ L$.
The corresponding wave functions are localized at a length scale
$\sim L(k_F^2\tilde a \xi)^{-1/4}$.
Such localization effect should result, in particular,
in the suppression of the heat transport in the $z-$ direction.

We thank D. Ryzhov, V. Pozdnyakova, and I. Shereshevskii for help
with computer codes, and N. Kopnin and G. Volovik for stimulating
discussions. This work was supported, in part, by Russian Foundation
for Basic Research, by Program ``Quantum Macrophysics'' of RAS, and by Russian
Science Support and ``Dynasty'' Foundations. ASM acknowledges the support
by the Academy of Finland.

 \end{document}